\documentclass[prb,showpacs,amsmath]{revtex4}
\begin{document}
\title{Boundary Conditions in an Electric Current 
Contact}
\author{O.~Yu.~Titov}
\email{oleg.titov@aleph-tec.com}
\affiliation{CICATA---IPN, Av.~Jos\'{e} Siurob 10, Col.~Alameda,\\
76040 Quer\'{e}taro, Qro., M\'{e}xico}

\author{J.~Giraldo}
\email{jgiraldo@ciencias.unal.edu.co}
\affiliation{Grupo de F\'{\i}sica de la Materia Condensada,\\
Universidad Nacional de Colombia, A.A. 60739, Bogot\'{a}, Colombia}

\author{Yu.~G.~Gurevich}
\email{gurevich@fis.cinvestav.mx}
\affiliation{Depto.~de~F\'{\i}sica, CINVESTAV---IPN,\\
Apdo.~Postal 14--740, 07000 M\'{e}xico, D.F., M\'{e}xico}

\begin{abstract}
In most electronic devices, electric current of both 
types (electrons and holes) flows
through a junction. 
Usually the boundary conditions 
have been formulated 
exclusively for open circuit. 
The boundary
conditions proposed here bypass this limitation by the first time, 
as far as
we are aware. Besides, these new boundary conditions correctly describe current 
flow in a circuit, i.e., closed circuit conditions, which are the usual 
operation conditions for electronic devices and for the measurement 
of many transport properties. 
We also have generalized the case 
(as much as it is possible in a classical treatment), 
so
self-consistent boundary conditions
to describe current-flow through a contact between two arbitrary
conducting
media are developed in the present work. 
These boundary conditions take into account a
recently developed theory: influence of temperature space
inhomogeneity due to the interfaces and quasi-particles
temperature-mismatch on
thermo-generation and recombination.
They also take into account
surface
resistance, surface recombination rates and possible temperature 
discontinuities
at the interface due to finite surface thermo-conductivity. 
The temperature difference between current-carriers and phonon
subsystems is also included in this approach. 
\end{abstract}

\maketitle

This work addresses the problem of defining appropriate boundary
conditions (BCs) in a closed electronic circuit of any type in 
the simplest general case. It is a continuation and generalization
of previous 
studies published elsewhere.\cite{g1,g2,g3,g4,g5}
They have been undertaken to clarify different processes
arising in connection with real working conditions of electronic 
devices in general,\cite{Singh,Sze} and in the search for 
applications to design 
new optoelectronic and thermoelectric devices.\cite{Aguado,DiSalvo,Fujiwara}

In spite of great advances in the field of contemporary electronics, 
the effects of any boundary between two arbitrary materials have 
been studied, in practice, only for open circuits. 
The presence of
both charge carriers (like electrons and holes in semiconducting
materials; Cooper pairs in superconductors; etc...) in 
circuits containing heterojunctions, makes a big difference when the
circuit is closed. 
Additionally, an intrinsic nonlinear effect due to the temperature
difference between electrons and holes, from one
side, and phonons, from the other (hot electrons), is usually
forgotten (see, for example Ref.~\onlinecite{Tautc,Anselm,Ioffe}). 
This effect has been studied in Ref.~\onlinecite{g5}.

Actually, one can mention very few
works handling closed circuits and non-equilibrium 
carriers (see Ref.~\onlinecite{Balmush} and reference therein). 
In more recent works\cite{g1,g2,g4} it has been shown that it is
compulsory to take into account non-equilibrium carriers in
bipolar and $p$-type semiconductors even 
in a linear approximation by electric field; furthermore, 
recombination starts to play
an important role 
even in the linear regime
and cannot be neglected.\cite{g3,g4,g5}

In practice, people use zero-current BCs 
to  define the functionality of solid state
electronic devices in general.\cite{Singh,Sze} This is an ideal assumption
that works well in many cases, particularly 
when there is no
different kind of charge and heat carriers and interfaces
involved. For a proper description of   
real performance conditions in heterojunctions, it should not 
be ignored that all devices are working
in a mode such that electric current flows through the ends of
a semiconducting structure and in a whole closed circuit, completed with
another semiconductor, a normal metal or a superconducting material.
Typically the semiconductors have both charge carriers, electrons and
holes, and important recombination processes take place at the ends.
All these facts demand a more stringent choice of BCs, and in
particular of non-vanishing current BCs at the borders between
different elements of a circuit to setup the correct
working conditions of solid state  electronic devices in general.
These BCs become more important nowdays when accurate measurements of 
current to the level of holes and electrons are feasable.\cite{Fujiwara}

In what follows, general BCs at the contact between two arbitrary 
conducting media will be obtained.  Next we  exemplify the usual
case of a metal-semiconductor boundary. Finally we comment our results. 

BCs (see, for example, Ref.~\onlinecite{Landau})
can be obtained from the transport
equations for non-equilibrium  carriers.\cite{Bonch}
In the static case they look like:
\begin{equation}
\operatorname{div} j_n = eR_n,\qquad 
\operatorname{div} j_p = -eR_p.
\end{equation}
Here $R_{n,p}$ are electron and hole recombination rates; 
\begin{equation}
j_n = \sigma_n
\left[
-
\frac{d\tilde\varphi_n}{dx} 
-
\alpha_n\frac{dT_n}{dx}
\right],\qquad
j_p = \sigma_p
\left[
-
\frac{d\tilde\varphi_p}{dx} 
%- \frac{T^{*}}{ep_0}\frac{d\delta p}{dx}
-
\alpha_p\frac{dT_p}{dx}
\right],
\end{equation}
are electron and hole electric currents; 
$\sigma_{n,p}$---electron and hole conductivity; 
\begin{equation}
\tilde\varphi_{n,p} = \varphi \mp \frac{\mu_{n,p}}{e}
\end{equation}
are the electrochemical potentials 
(for exhaustive discussion see Ref.~\onlinecite{gm}),
where $\varphi$ is the electric potential and $\mu_{n,p}$ 
are electron and hole chemical potentials; the latters are related by
$\mu_p = -\epsilon_g - \mu_n$ ($\epsilon_g$ is the semiconductor
band gap) only under thermodynamic equilibrium conditions 
(for exhaustive discussion see Ref.~\onlinecite{g3});
($-e$) is the electron charge
($e>0$); $\alpha_n$ and $\alpha_p$, electron and hole power coefficients; 
$T_n$ and $T_p$---electron and hole temperatures 
(for exhaustive discussion see Ref.~\onlinecite{gm}).
The recombination rates are usually written out in the following 
form:~\cite{Neamen}
\begin{equation}
R_n = \frac{\delta n}{\tau_n},
\qquad
R_p = \frac{\delta p}{\tau_p}. 
\end{equation}
Here $\tau_{n,p}$ are electron and hole lifetimes;
$\delta n = n - n_0$ and $\delta p = p - p_0$
are electron and hole non-equilibrium concentrations; 
$n$ and $p$ are the full concentrations; 
$n_0$ and $p_0$ are the corresponding equilibrium values; 
$\delta n$ and $\delta p$
are related to fluctuations in the
chemical potentials $\delta\mu_{n,p}$.\cite{g4} 
Notice that these and the Fermi quasilevels   
$\delta\tilde\varphi_{n,p}(x) = \delta\varphi(x) \mp
(1/e)\delta\mu_{n,p}(x)$  are inhomogeneous across a boundary.

It follows from Maxwell's equations that 
$\operatorname{div} (j_n + j_p) = 0$ in the steady state, so
that taking into account Eq.~(4) one obtains from  Eq.~(1) the unphysical
condition: $(\delta n/\tau_n) = (\delta p/\tau_p)$. Gurevich
{\it et al.} (in Ref.~\onlinecite{g4}) have shown that Eq.~(4) is 
thermodynamically incorrect; moreover, Eqs.~(4) should always be replaced 
by the following relationship (according to Ref.~\onlinecite{g3}):
\begin{equation}
R_n = R_p = R,\qquad
R = \frac{\delta n}{\tau_n} + \frac{\delta p}{\tau_p}.
\end{equation}

The system of Eqs.~(1) is incomplete as we have three unknown functions
$\delta n(x)$, $\delta p(x)$, and $\varphi(x)$.  Poisson equation might be
used to complete it. For simplicity, we will assume that all
characteristic  lengths are much bigger than the Debye's radius,\cite{Silin}
so that:
\begin{equation}
\delta n(x) = \delta p(x)
\end{equation}
and Poisson equation becomes unnecessary. 

Let us assume that the boundary between media 1 and 2 lies at $x = 0$. 
Integrating Eqs.~(1) with $x$ in a short range from $x = -\delta$ to
$x = +\delta$ and taking the limit $\delta \to 0$ one obtains: 
\begin{equation}
j_n (+0) - j_n (-0) = +eR_s,\qquad
j_p (+0) - j_p (-0) = -eR_s.
\end{equation}
Here 
\begin{equation}
R_s = \lim_{\delta \to 0} \int_{-\delta}^{+\delta}R(x)\,dx
\end{equation}
is the surface recombination rate. Making use of Eq.~(5) it can be
rewritten as:\cite{g3} 
\begin{equation}
R_s = S_n \delta n + S_p \delta p, 
\end{equation}
or, with the help of Eq.~(6), 
\begin{equation}
R_s = S \delta n,\qquad
S = S_n + S_p.
\end{equation}

Notice that the recombination rate $R$ in Eq.~(8) is $x$ dependent.
Each one of the contacting media changes its properties in a distance
close to the Debye's radius and becomes  inhomogeneous in that region. 
This explains the $x$ dependence of $R$ 
in Eq.~(8).

Let us now integrate Eq.~(1) with  $dx$ from $\varepsilon$ to $\delta$,
and with $d\varepsilon$  from $-\delta$ to
$+\delta$. One obtains: 
\begin{equation}
\begin{aligned}
j_n(+0) = & \sigma_{n}^{s}
\left\{
\tilde\varphi_n(-0) - \tilde\varphi_n(+0) 
   - \alpha_n^s
\left[
T_n(+0) - T_n(-0)
\right]
\right\}
+ eR_s^{n+}\\
j_p(+0) = & \sigma_{p}^{s}
\left\{
\tilde\varphi_p(-0) - \tilde\varphi_p(+0) 
   - \alpha_p^s
\left[
T_p(+0) - T_p(-0)
\right]
\right\}
- eR_s^{p+}.
\end{aligned}
\end{equation}
Here $\sigma_n^s$ and $\sigma_p^s$ 
are electron and hole surface conductivity: 
\begin{equation}
(\sigma_{n,p}^s)^{-1}= 
\lim_{\delta \to 0} \int_{-\delta}^{+\delta}
(\sigma_{n,p})^{-1}(\xi)d\xi,
\end{equation}
$\alpha^s_{n,p}$ are  thermopower surface coefficients:
\begin{equation}
\alpha^s_{n,p} = \frac{1}{T(\delta)-T(-\delta)}
\lim_{\delta \to 0} \int_{T(-\delta)}^{T(\delta)}
\alpha_{n,p}(T)dT, 
\end{equation}
$R^{n+,p+}_s$ are surface recombination rates for electrons 
and holes to the right of the boundary $x=0$: 
\begin{equation}
R^{n+}_s 
= \sigma_n^s \lim_{\delta \to 0}
\int_{-\delta}^{+\delta}d\varepsilon{\sigma_{n}}^{-1}(\varepsilon)
\int_{\varepsilon}^{\delta}dx R(x)\qquad
R^{p+}_s = \sigma_p^s\lim_{\delta \to 0}
\int_{-\delta}^{+\delta}d\varepsilon{\sigma_{p}}^{-1}(\varepsilon)
\int_{\varepsilon}^{\delta}dx R(x).
\end{equation}
Using the same procedure, the following BCs are obtained for the left:
\begin{equation}
\begin{aligned}
j_n(-0) = & 
\sigma_n^s
\left\{
\tilde\varphi_n(-0) - \tilde\varphi_n(+0) 
- \alpha_n^s
\left[
T_n(+0) - T_n(-0)
\right]
\right\}
- eR_s^{n-}\\
j_p(-0) = &
\sigma_p^s
\left\{
\tilde\varphi_p(-0) - \tilde\varphi_p(+0) 
- \alpha_p^s
\left[
T_p(+0) - T_p(-0)
\right]
\right\}
+ eR_s^{p-}.
\end{aligned}
\end{equation}
$R^{n-,p-}_s$ are surface recombination rates for electrons 
and holes to the left of the boundary $x=0$: 
\begin{equation}
R^{n-}_s = 
\sigma_n^s\lim_{\delta \to 0}
\int_{-\delta}^{+\delta}d\varepsilon{\sigma_n}^{-1}(\varepsilon)
\int_{-\delta}^{\varepsilon}
R(x)\,dx\qquad
R^{p-}_s =
 \sigma_p^s\lim_{\delta \to 0}
\int_{-\delta}^{+\delta}d\varepsilon{\sigma_p}^{-1}(\varepsilon)
\int_{-\delta}^{\varepsilon}
R(x)\,dx.
\end{equation}

From Eqs.~(8), (14), and (16) it becomes evident that
\begin{equation}
R^{n+}_s + R^{n-}_s = R^{p+}_s + R^{p-}_s = R_s
\end{equation}
therefore BCs  (7), (11), and (15) are not independent and
only two of them should be used.

One should base particular decision of which BCs to use on experimental
setup or physical sense of the problem to solve. Besides, this decision
depends a lot on particular properties of the contact (Schottky barrier,
$p$-$n$-junction, $n^+$-$n$-contact, etc...).  We will demonstrate below
how to choose the correct BCs for a metal-semiconductor junction.

We would like to emphasize two important facts. Firstly, it is not enough
to define general surface recombination rates for the correct definition
of transport effects on contact (this follows from Eqs.~(11)  and~(15)).
One has to use particular surface recombination rates for every type of
current carriers (electrons, holes, etc.) defined on both sides of the
contact. As far as we are aware, this has not been taken into account
previously by anybody.

Secondly, we have used electron's and hole's Fermi quasilevels in
Eqs.~(11) and~(15). One assumes that the electron's Fermi quasilevel is
measured from the bottom of the conduction band, and that the hole's
quasilevel is measured from the top of the valence band. Usually the
position of the top of the valence band and the bottom of the conduction
band are different in heterocontacts. That is why the reference (zero)
points for the Fermi quasilevels are different at $x = +0$ and $x = -0$.

Next we will describe how to take into account the above mentioned
differences in a metal--semiconductor boundary.

It is very common to inject electric current into semiconductor samples
through metal contacts. These contacts have an interesting peculiarity:
semiconductors can be characterized by the presence of both electrons and
holes, while metals have only electrons as current carriers.

Let us take into account this fact in Eqs.~(7), (11), and (15).  We assume
that we have a metal to the left of $x = 0$ and a semiconductor to the
right. Therefore we have that $j_p^m(-0) = 0$, so that Eq.~(7)  leads to
$j_p^s(+0) = - eR_s$ (the subscript indicates the corresponding media;
here ``s'' stands for semiconductor and ``m'' for metal).

We can simplify Eq.~(15) due to the absence of surface recombination
in a metal (it has only electrons as charge carriers): 
$R_s^{n-} = R_s^{p-} = 0$.
Therefore, from Eq.~(17) it follows that:
\begin{equation}
R_s = R_s^{n+} = R_s^{p+}.
\end{equation}
There are no positive charge carriers in metals, which means 
that electron current in the metal,  $j_n^m(-0)$, should be  
equal to the whole current $j_0$.
In this case, the BCs (15) reduce to: 
\begin{equation}
j_0 = \sigma_n^s
\left\{
\varphi_m - \varphi_s(0) - \frac{\mu_m}{e} + \frac{\mu_s(0)}{e} 
+\frac{\Delta\varepsilon_c}{e}
-
\alpha_n^s
\left[
T_n^s(0) - T_n^m(0)
\right]
\right\}.
\end{equation}
Notice that the electrical and the chemical potential of the 
metal have not changed.
Let us remind that $\mu_m$ and $\mu_s$ are calculated
from the bottom of the  
conduction band of each material. This leads to the additional 
term in Eq.~(19) $(\sigma_n^s/e)\Delta\varepsilon_c$, where
$\Delta\varepsilon_c$ is the  distance between the metal and
the semiconductor conduction bands.

It follows from the second equation of system (15) that 
$\sigma_p^s = 0$. 
This is to be expected since the absence of holes in a  metal 
means the absence of holes surface conductivity. 

Summarizing all the above mentioned facts, we can write 
Eq.~(11) as
\begin{equation}
j_n^s(0) =  \sigma_n^s 
\left\{
\varphi_m - \varphi_s(0) - \frac{\mu_m}{e} + \frac{\mu_s(0)}{e} 
+\frac{\Delta\varepsilon_c}{e}
- \alpha_n^s
\left[
T_n^s(0) - T_n^m(0)
\right]
\right\}
+ eR_s,\qquad
j_p^s(0) =  -eR_s.
\end{equation}
The last condition in (20) is equal to the second condition 
in (7). The first
equation in (7) can be rewritten as 
\begin{equation}
j_n^s(0) - j_0 = eR_s.
\end{equation}

Notice that the first equation (20) follows from Eqs.(19) and 
(21). All this yields the new BCs for a metal-semiconductor
junction which are now formed by Eq.~(19), Eq.~(21), and the second
of equations (20). 

In conclusion, we have formulated general BCs, corresponding to current
flow through the boundary between two conducting media. These conditions
consider possible jumps of the electron's and hole's Fermi quasilevels and
of the electric potentials at the boundary; they also take into account
the surface recombination rates. The general procedure has been applied to
a usual contact, namely the metal-semiconductor boundary, but the method
is valid for all types of contacts ($n^+$-$n$, $p^+$-$p$, $p$-$n$,
Schottky barrier, etc...) between different conducting materials.

This work was partially supported by CONACyT, Mexico. JG has received
support from DINAIN (Universidad Nacional de Colombia), Colombia.

\end{document}